# The Vacuum Energy in a Condensate Model for Spacetime

Gerorge Chapline
Lawrence Livermore National Laboratory
Livermore, CA 94507

**Abstract**

It is shown that a simple model for 4-dimensional quantum gravity based on a 3-dimensional generalization of anyon superconductivity can be regarded as a discrete form of Polyakov's string theory. This suggests that there is a universal negative pressure that is on the order of the string tension divided by the square of the Robertson-Walker scale factor. This is in accord with recent observations of the brightness of distant supernovae, which suggest that at the present time there is a vacuum energy whose magnitude is close to the mass density of an Einstein de Sitter universe.

## 1.Introduction

Recent observations [1] of Type I supernovae at cosmologically significant redshifts have tended to confirm old suspicions [2] that the average density of matter is smaller than that required for a flat ($\Omega=1$) universe. In addition, these observations suggest that there is at the present time a positive vacuum energy whose magnitude is comparable to the average matter density. This last result is very surprising from the point of view of theoretical expectations based on either conventional 4-dimensional quantum field theories or the types of superstring models of elementary particles that have been intensively investigated during the past few years. On the other hand, the author proposed some time ago [3] a model for the ground state of quantum gravity whose structure is rather different from that of the ground states for either 4-dimensional field theories or superstring models. This model requires that on length scales large compared to the Planck length but small compared to astronomical length scales four dimensional space-time is geometrically flat. In this paper we will address the question of the vacuum energy in this model.

The fundamental degrees of freedom in the model of ref.3 are just those of a supersymmetric $SU(\infty)$ gauge theory defined on a 2-dimensional surface, but can be equivalently described as the moduli of self-dual and anti-self-dual ALE spaces [4]. These moduli can also be thought of [4] as the "positions" of quasi-particles moving in 3-

dimensions. In the limit where self-dual and anti-self-dual quasi-particles are paired so that their positions closely coincide, this model apparently provides a quantum description of classical flat spacetime, while low lying excitations of the ground state may include both supergravity as well as phenomenological matter and gauge fields [5]. In this note we would like to point out that the emergence of gravity-like classical fields in our quantum gravity model is actually closely related to the way that 2-dimensional conformal field theory emerges from a pairing of world sheet vortices and anti-vortices in a discrete version of Polyakov's string theory [6]. Furthermore just as the pairing of world sheet vortices and anti-vortices leads to a positive string tension it is to be expected that the model of ref.3 will lead to a ground state for quantum gravity with a positive vacuum energy. In fact because the model of ref.3 also requires that on average spacetime be flat, in the absence of matter this vacuum energy would have the critical value $\rho_c = 3H^2/8\pi G$, where H is the Hubble constant. In the present epoch where the density of ordinary matter is not completely negligible compared to the critical density, the vacuum energy density would have a value somewhat less than the critical density. However, in contrast with previously considered cosmological models the vacuum energy for the model of ref.3 does not correspond to a cosmological constant in the usual sense, since the vacuum energy is inversely proportional to the square of the cosmological distance scale, and therefore decreases with time. As will be seen this leads to a new type of cosmological model that resembles in some ways an open Friedman universe with no cosmological constant. In the following section we explain why the condensate model of ref. 3 gives rise to a positive vacuum energy that is inversely proportional to the square of the scale factor for the universe, while in section 3 we discuss possible cosmological signatures for this type of time varying vacuum energy.

**2. Why is there a universal negative pressure?**

A negative pressure in a material medium can arise if there are attractive interactions between the constituent particles or if the entropy of the medium decreases as its volume increases. In the case of a superfluid the Bose statistics of the contituent particles give rise to an effective attractive interaction [7], and therefore a superfluid wll have an effective negative pressure associated with its superfluidity. We will argue here that the gound state of ref.3 has a negative pressure which is related to this latter effect. In particular, we will argue that the peculiar holonomy properties of the solitonic quasi-particles whose moduli describe the microscopic degrees of freedom of the vacuum state give rise to an effective attractive interaction that is very similar to the string tension in conformal field theory. These holonomy properties are reminiscent of the fractional statistics of 2-dimensional anyons and reflect the 2-dimensional nature of the underlying theory. Furthermore, just as the fractional statistics of the quasi-particles in the fractional quantum Hall effect ground state lead to a special type of quantum long range order [8], we expect that the holonomy properties of the quasi-particles in our ground state imply the existence of a quantum long range order; which was interpreted in ref.3 to mean the appearence of a classical spacetime.

The ground state wavefunction introduced in ref.3 has the form

$$\Psi = f(\bar{z}) \prod_{k>j}^{\infty} \left[ \frac{R_{jk} + U_{jk}}{R_{jk} - U_{jk}} \right]^{1/2}, \tag{1}$$

where $R_{jk}^2 = U_{jk}^2 + 4(z_j - z_k)(\bar{z}_j - \bar{z}_k)$, $U_{jk} = U_j - U_k$, and $f(\bar{z})$ is an entire function of the $\bar{z}_j$. The product on the right hand side of (1) replaces the $\prod_{k>j} |z_j - z_k|^{-\theta/\pi}$ factor in Laughlin-type wavefunctions [9]. Like the Laughlin anyons, which obey fractional statistics determined by the parameter θ, the quasi-particles (referred to as "chirons" in ref.4) described by the wavefunction (1) have interesting holonomy properties. The complex phase or "action" associated with a chiron whose 3-dimensional position is parameterized by z=x+iy and U will be given by

$$\phi = \frac{1}{2} \sum_j \ln \frac{R_j + U - U_j}{R_j - U + U_j}, \tag{2}$$

where $R_j^2 = (U - U_j)^2 + 4(z - z_j)(\bar{z} - \bar{z}_j)$. The change in phase moving around $z = z_k$ is

$$\oint_k \phi_z dz = \frac{1}{2} \sum_i \oint_k \frac{U_{ji}}{R_i} \frac{dz}{z - z_i} = i\pi. \tag{3}$$

Thus for cyclic changes in z chirons behave like anyons with 1/2 fractional statistics. It should be noted though that the holonomy is non-trivial only if $U_{jk} \neq 0$, so that the physical effects of the non-trivial holonomy exist only in 3 spatial dimensions. We will now show that one of these physical effects is that the ground state (1) has a negative pressure.

Despite the fact that the ground state wavefunction (1) is intended to describe a classical 4-dimensional spacetime, the product on the right hand side of (1) has a remarkable connection with Polyakov's string action. It was shown some time ago [6] that a 2-dimensional conformally invariant action of the Polyakov form can be interpreted as arising from the condensation of 2-dimensional quantized vortices. This condensation of 2-dimensional vortices is essentially the same as the Kosterlitz-Thouless (KT) transition [10] in the XY model for a 2-dimensional ferromagnet. The KT transition in the XY model consists of a condensation of vortex and anti-vortex configurations of planar spins into bound state pairs. Each XY spin is described by an angle $\Theta$; and the vortex configurations implicated in the KT transition have the form:

$$\Theta(z) = m_i \, \text{Im} \, \ln(z-z_i), \qquad (4)$$

where the integer $m_i$ is the quantized circulation of the vortex (or anti-vortex if $m_i$ is negative) located at $z_i$. On the high temperature side of the KT transition the XY model is equivalent to a 2-dimensional gas of vortices with Coulomb-like interactions, while on the low temperature side of the transition the model becomes a scale invariant theory of massless spin waves [11,12]. In ref. 6 it was shown that in a very similar way to the XY model the 2-dimensional conformally invariant action corresponding to a "physical" string theory corresponds to the low temperature side of a KT-like transition in which world sheet vortices and anti-vortices similar in form to the KT vortices (4) are bound together. What is particularly important for us is that this derivation of a continuum conformally invariant string action is based on a discrete version of the Polyakov string action which, like the XY spin model, interpolates between a 2-dimensional conformally invariant field theory at low temperatures and a theory of interacting quantized vortices at high temperatures. We will now show that this result provides us with a remarkable way of looking at our quantum gravity model.

Comparing the KT vortices (4) with the expression (2) for the complex phase of a chiron one eventually realizes that there is a remarkable similarity between the two expressions. Indeed it is an elementary identity that the right hand side of (2) can be rewritten in the form

$$\phi = \sum_i \tanh^{-1} \left( \frac{U - U_i}{R_i} \right), \qquad (5)$$

which except for the appearance of a hyperbolic in the place of a circular tangent function is a sum of terms of the same form as (4). In ref.6 it was noted that if we identify $\Theta$ with a periodic string coordinate whose period is L, then the usual Polyakov path integral evaluated for a sum of configurations of the form (4) yields (neglecting a divergent constant) an expression of the form:

$$\sum_{m_r, z_i} \exp - \frac{TL^2}{4\pi} \left[ \sum_{r \neq s} m_r m_s \ln \frac{1}{|z_r - z_s|} \right], \qquad (6)$$

where the sum over vortices is restricted to zero total circulation and T is the usual string tension. It is worth noting that this expression can be interpreted as the partition function for a 2-dimensional gas of charged particles interacting via a Coulomb potential. In a similar way substituting the chiron phase (5) into the Polyakov string action yields an expression with exactly the same form as (6) except that $\ln[1/|z_r - z_s|]$ is replaced by $\ln[R_{rs}/|z_r - z_s|]$ and $TL^2/4\pi = \pi$. The KT transition occurs for $TL^2/4\pi = 2$. Therefore evaluating the Polyakov action for the chiron configurations (6) leads to an expression which resembles the partition function of a Coulomb gas just below the KT transition. It can be shown [11,12] that at low temperatures one can rewrite the partition function for a 2-dimensional Coulomb gas in the form:

$$\prod_j \int_{-\infty}^{\infty} \frac{d\vartheta(j)}{2\pi} \exp\left(-\frac{K}{2} \sum_{i \neq j} [\vartheta(i) - \vartheta(j)]^2\right), \tag{7}$$

where K is a constant. If we identify $K = TL^2/4\pi^2$ and $X = L\vartheta$, then in the continuum limit expression (7) becomes the usual scale invariant string action.

Evidently the quantity $<U_i - U_j>$ serves as a length scale such that for physical values of $|z_r - z_s|$ less than $<U_i - U_j>$ the interaction between chirons resembles the interaction between vortices in the XY model. This means that in accordance with our previous speculations the long range order in our ground state will be closely related to the topological order [10] that appears on the low temperature side of the KT phase transition. Furthermore, since in our ground state $<|z_r - z_s|> \approx <U_i - U_j>$, it is clear that the effective action for our model ground state will resemble that for a string compactified on a circle with circunference $2\pi/\sqrt{T}$, and that this length scale will be on the order of the mean distance between chirons. A priori there is no reason to connect this length scale with the Planck length. However, if our model for quantum gravity proves succesful as a theory of classical gravity [see e.g. ref.5], we expect that it will also be possible to identify the Planck length as the inverse of the square of the mean distance between chirons. It is of course probably not accidental that such an identification would be consistent with the Scherk-Schwarz identification of the Planck constant with a string tension [13].

We have found that there is a 1-dimensional string-like tension associated with our model ground state for quantum gravity. There is naturally no preferred orientation for the 2-dimensional surfaces of the underlying theory, so that when averaged over all possible orientations for these surfaces the 1-dimensional tension becomes a negative pressure. However this negative pressure will be smaller than the string tension by a factor on the order of the square of the length scale for the universe. Thus we find a value for the vacuum energy that is enormously smaller than that found in conventional field theories.

## 3. A Pseudo-Open Universe?

For many years now a dichotomy has existed between the bulk of astronomical and cosmochemical observations, which seemed to favor a low density open universe [2], and aesthetic and theoretical prejudice, which has favored a flat ($\Omega$=1) universe. This dichotomy has given rise to a minor industry : the search for "dark matter" on cosmological scales. On the other hand as mentioned in the introduction recent observations of distant supernovae have suggested that there may be a non-zero vacuum energy, which might turn out to be the dark matter needed to make the universe flat. As we have just seen there are some theoretical reasons for believing that indeed from the point of view of quantum gravity not only is the universe geometrically flat, but the vacuum energy is non-zero and (in Planck units) on the order of the inverse square of the distance scale parameter for a Robertson-Walker metric. This would imply that the cosmological "constant" is not time independent. Of course, the possibility of a time varying vacuum energy has always existed, and a priori there is no obvious reason why cosmological models with a vacuum energy varying inversely with the square of the scale parameter should be excluded.

The scale factor R for a Robertson-Walker universe containing cold matter and with a cosmological constant satisfies the "energy" equation [14] :

$$H^2 = \frac{8\pi}{3} G\rho + \frac{\Lambda}{3} - \frac{\kappa c^2}{R^2}, \qquad (8)$$

where $H = \dot{R}/R$ is the Hubble constant at time t and $\kappa = \pm 1$ or 0 specifies whether the universe is closed, open, or flat. For a flat universe where $\Lambda = \lambda/R^2$ we have

$$H^2 = \frac{8\pi}{3} G\rho + \frac{\lambda}{3R^2}. \qquad (9)$$

This equation resembles the equation for the Hubble constant for an open Friedman universe, except that in an open Friedman universe the Einstein equations would require that $\lambda$=3. However eq. (9) can be integrated as it stands for any value of $\lambda$, yielding the following expressions for the Hubble constant and deceleration parameter q as a function of R :

$$H = \frac{\sqrt{m}}{R^{3/2}} (1 + \frac{\lambda R}{3m})^{1/2} \qquad (10)$$

$$q = \frac{\Omega_m}{2}, \qquad (11)$$

where m=$(8\pi G\rho/3)R^3$ and $\Omega_m$ is the ratio of the density of ordinary matter to the critical density. At early times when $\Omega_\Lambda \ll \Omega_m$, H and q have values characteristic of an Einstein-de Sitter universe, while at late times when $\Omega_\Lambda \gg \Omega_m$, H and q behave in a manner similar to that of an open Friedman universe. The age of the universe in our cosmological model will be given by

$$t_0 = \frac{1}{H_0} \left[ 1 + \frac{1}{u^2} - \frac{\sqrt{1+u^2}}{u^3} \ln(u + \sqrt{1+u^2}) \right], \qquad (12)$$

where $u^2 = \frac{\lambda R}{3m}$. The expressions (11) and (12) have exactly the same form as for an open Friedman universe with no cosmological constant, except that for the Friedman case $u^2 =$ R/m. At early times when $\Omega_\Lambda \ll \Omega_m$ $t_0$ is close to the Einstein-de Sitter value $\frac{2}{3H_0}$, while at late times $t_0$ approaches the Milne value $1/H_0$. Similarly at early times the deceleration parameter q is close to the Einstein-de Sitter value 0.5, while at late times q approaches zero. Thus the deceleration parameter and the relationship between the age of the universe and Hubble constant behave in a manner quite similar to that in an open Friedman universe with no cosmological constant. It should be noted, though, that the predicted late time behaivor of these quantities in our model with a time varying vacuum energy is quite different from what would be expected in any Robertson-Walker model with a time independent cosmological constant. For such models the deceleration parameter q approaches -1 and the Hubble constant approaches a constant.

If one could accurately measure q or $t_0$ for the present epoch, it might be possible to distinguish between a flat universe with a time independent cosmological constant and one with a cosmological constant varying as $1/R^2$. Unfortunately up to the present time it has not been possible to directly measure these quantities with any acuracy. At present the most reliable information regarding cosmological parameters is provided by the recent measurements of the apparent brightness of distant supernovae. In our flat universe model with a vacuum energy varying as $R^{-2}$ the luminosity distance $d_L$ as a function of redshift will be given by

$$d_L = \frac{1}{H_0}(1+z)\,\Omega_\Lambda^{-1/2}\left[\cosh^{-1}(2u_0^2+1) - \cosh^{-1}(\frac{2u_0^2}{z+1}+1)\right]. \qquad (13)$$

The dynamics and X-ray emissions of galaxy clusters suggests [15] that at the present time $\Omega_m \approx 0.2$, which together with our requirement for flatness implies that at the present time

$\Omega_\Lambda \approx 0.8$. For the most distant supernovae observed by Perlmutter, et. al. [1] $z \approx 0.8$, in which case eq.12 using $\Omega_\Lambda \approx 0.8$ and $u_o^2 \approx 4$ predicts that $d_L = 1.18 H_0^{-1}$. By way of comparison in an Einstein-de Sitter universe one would have $d_L = 1.04 H_0^{-1}$. Our prediction for the brightness of supernovae with $z \approx 0.8$ is consistent with the observations, but unfortunately it is also close to what would be predicted by a model with a time independent $\Lambda$.

Observations of supernovae at greater redshifts might eventually allow one to discriminate between cosmological models with constant $\Lambda$ and our model where the vacuum energy varies as $R^{-2}$. However a better discriminate may perhaps be provided by measurements of the probability of gravitational lensing of distant quasars by intervening galaxies. It has been known for some time that the statistical probability of gravitational lensing of distant quasars is sensitive to the values of cosmological parameters [16]. In particular flat universe models with a time independent cosmological constant generally predict significantly larger probabilities for the gravitational lensing of large redshift quasars than models with no cosmological constant [17]. Modeling the lensing galaxies as point masses, and using the Press-Gunn formula for the gravitational lensing cross-section of a point mass [16], we obtain the following approximate formula for the lensing optical depth as a function of quasar redshift $z_Q$:

$$\tau = \frac{1}{2} \Omega_L \frac{z_Q^2}{2 + z_Q}. \tag{14}$$

This formula is identical with the formula for the gravitational lensing optical depth due to point masses in a low density open universe previously obtained by Turner, Ostriker, and Gott [16]. Although there is considerable uncertainty as to what would be a reasonable value for $\Omega_L$, the formula (14) appears to be at least consistent with the observed frequency of lensed quasars. In contrast, the consistency of a time independent cosmological constant with these observations appears at the present time to be rather problematic. Of course, we hope this issue will be resolved in the near future since such clarification could provide direct support for the novel type of cosmological model that our simple model for quantum gravity seems to imply.

**Conclusion**

We have shown that in a simple model for the ground state of quantum gravity, there is a very nice explanation for why there might be a positive vacuum energy whose value at the present epoch is close to the Einstein-de Sitter critical density. This explanation connects 4-dimensional quantum gravity with Polyakov's string theory in a heretofore unsuspected way. In particular, 4-dimensional quantum gravity appears as a more fundamental "topological" version of Polyakov's theory, and the vacuum energy is directly related to the string tension. The effective string tension in our theory arises from an attraction between topological solitons in the ground state that is closely related to the effective attraction between quasi-particles that responsible for the Einstein-Bose condensation observed in liquid helium and other quantum fluids. This connection of the ground state for quantum gravity with Bose superfluids is also noteworthy because it leads to a simple resolution of the black hole information puzzle [18]. Finally we note with some anticipation that our explanation for a positive vacuum energy leads to cosmological predictions that may be testable in the near future.

**References**

1. S. Perlmutter et. al., Discovery of a supernovae explosion at half the age of the universe. Nature 391 (1998), 51.
2. J. R. Gott , J. E. Gunn, D. Schramm, and B. M. Tinsley, An unbound universe?. Ap. J. 194 (1974), 543.
3. G. Chapline, A quantum model for space-time. Mod. Phys. Lett. A7 (1992), 1959; Anyons and coherent states for gravitons. Proc. XXI International Conference on Differential Geometric Methods in Theoretical Physics, ed. C.N. Yang, M. L. Ge,and X. W. Zhou (World Scientific 1993).
4. G. Chapline and K. Yamagishi, A three-dimensional generalization of anyon superconductivity, Phys. Rev. Lett. 66 (1991) 3046.
5. G. Chapline, A unique theory of gravity and matter, hep-th /9802180.
6. G. Chapline and F. Klinkhamer, Vortices in high temperature string theory. Mod. Phys. Lett. A4 (1989) 1063
7.. G. Chapline, Theory of the superfluid transition in liquid helium. Phys. Rev. A3 (1971) 1671l
8. S. M. Girvin and A. H. MacDonald, Off-diagonal long-range order, oblique confinement, and the fractional quantum hall effect. Phys. Rev. Lett. 58 (1987) 1252.
9. R. B. Laughlin, Elementary theory: the incompressible quantum fluid. The Quantum Hall Effect, eds. R. Prange and S. M. Girvin (Springer-Verlag).

10. J. M. Kosterlitz and D. J. Thouless, Ordering, metastability, and phase transitions in two-dimensional systems. J. Physics C6 (1973) 1181.

11. J. V. Jose, et. al., Renormalization, vortices, and symmetry-breaking perturbations in the two-dimensional planar model. Phys. Rev. B16 (1977) 1217.

12. R. Savit, Vortices and the low-temperature structure of the xy model. Phys. Rev. B17 (1978) 1340.

13. J. Scherk and J. H. Schwarz, Nucl. Phys. B81 (1974) 118.

14. S. Hawking and G. F. R. Ellis, The Large Scale Structure of the Universe (Cambridge University Press 1973).

15. S. D. M. White, et. al., The baryon content of galaxy clusters: a challenge to cosmological orthodoxy. Nature 366 (1993) 429.

16. E. L.Turner, J. P. Ostriker, and J. R. Gott, The statistics of gravitational lenses. Ap. J. 284 (1984) 1.

17. E. L. Turner, Gravitational lensing limits on the cosmological constant in a flat universe. Ap J. 365 (1990) L43.

18. G. Chapline, The black hole information puzzle and evidence for a cosmological constant, hep-th /9807175.